\documentstyle[11pt]{article}
\newcounter{thnks}
\newcommand{\thankfn}[1]{\setcounter{thnks}{#1}$\mbox{}^{\fnsymbol{thnks}}$}
\begin{document}
\newcommand{\beq}{\begin{equation}}
\newcommand{\eeq}{\end{equation}}
\font\vbig=cmbx10 scaled\magstep3
\font\big=cmbx10 scaled\magstep1
\begin{titlepage}

\title{Optical Properties of Deep Ice at the South Pole: Absorption}
 
\author{The AMANDA Collaboration:\\
P. Askebjer\thanks{\it Stockholm University, Sweden},
S.W. Barwick\thanks{\it University of California, Irvine, USA},
L. Bergstr\"om\thankfn{1}, A. Bouchta\thankfn{1},
S. Carius\thanks{\it Uppsala University, Sweden},\\
E. Dalberg\thankfn{1}, K. Engel\thankfn{4},
B. Erlandsson\thankfn{1}, A. Goobar\thankfn{1}, 
L. Gray\thanks{\it University of Wisconsin, Madison, USA},
A. Hallgren\thankfn{3},\\
F. Halzen\thankfn{4}, 
H. Heukenkamp\thanks{DESY Zeuthen, Germany}
P.O. Hulth\thankfn{1},
S. Hundertmark\thankfn{5}
J. Jacobsen\thankfn{4}, \\
A. Karle\thankfn{5}
V. Kandhadai\thankfn{4}, I. Liubarsky\thankfn{4},
D. Lowder\thanks{\it University of California, Berkeley, USA},\\
T. Miller\thankfn{6}\thanks{\it
Currently at Bartol Research Institute, Delaware, USA}, P. Mock\thankfn{2},
R. Morse\thankfn{4},
R. Porrata\thankfn{2}, P.B. Price\thankfn{6},
A. Richards\thankfn{6},\\ H. Rubinstein\thankfn{3}, E. Schneider\thankfn{2},
Ch. Spiering\thankfn{5}, O. Streicher\thankfn{5}
Q. Sun\thankfn{1}, Th. Thon\thankfn{5},\\
S. Tilav\thankfn{4}, R. Wischnewski\thankfn{5}, Ch. Walck\thankfn{1}
\& G. Yodh\thankfn{2}\\
{\bf Corresponding author: L.\,Bergstr\"om }\\
e-mail: lbe@physto.se  }

\maketitle
\vskip .1cm
 \vskip .7cm
\begin{center}
{\bf Abstract}
\end{center}
\vskip .1cm
{\small We discuss recent measurements of the wavelength-dependent
absorption  coefficients in deep South Pole ice. The
method uses transit time distributions of  pulses from a
variable-frequency laser sent between emitters and receivers embedded
in the  ice. At depths of 800 to 1000 m scattering is dominated by
residual air bubbles, whereas  absorption occurs both in ice
itself and in insoluble impurities. The absorption
coefficient increases  approximately exponentially with wavelength in the
measured interval 410 to 610 nm. At  the shortest wavelength our value is about
a factor 20 below previous values obtained for laboratory ice and lake ice; with
increasing wavelength the discrepancy with previous measurements
decreases. At $\sim  415$ to
$\sim 500$ nm the experimental uncertainties
 are small enough for us to resolve an
extrinsic  contribution to absorption in ice: submicron dust particles
contribute by an amount that increases with depth and corresponds
well with the expected increase seen near the Last  Glacial Maximum in
Vostok and Dome C ice cores. The laser pulse method allows remote 
mapping of gross structure in dust concentration as a function of
depth in glacial ice.

}

\end{titlepage} 

\section{Introduction}

The AMANDA experiment (Antarctic Muon and Neutrino Detector Array)
is a U.S.-Germany-Sweden collaboration with the aim of deploying
a facility to detect neutrinos from astrophysical
sources (a ``neutrino telescope'') deep in the South Pole 
ice cap \cite{1a}.
High-energy muon neutrinos
will convert into charged muons through charged weak current interactions  
somewhere in the ice volume (or in the bedrock below the ice cap).
As the muons are very penetrating, and move at 
relativistic speed, they
will generate Cherenkov light along, typically, several hundred meters of track
length. By recording the arrival times of Cherenkov photons at several
locations in the ice volume, the direction of the muon track can be
reconstructed. Interesting events will be muons  coming from below
the horizon, since they can only have been generated by upcoming 
neutrinos (all other known particles would be stopped in the interior of 
the earth).

The AMANDA detector will eventually consist of many strings, each with  20--50 
optical modules containing photomultiplier tubes
(PMTs) lowered around 2 kilometers into the ice. Typical spacing between 
strings
and between modules in strings will be tens of meters. 

The performance
of the detector will depend crucially on the optical properties of the
ice. A first stage of the detector, deployed during
the 1993-94 austral summer, consists of four vertical strings each having
20 PMTs
spaced at 10 m distance. The strings form an $80\deg$ sector of a 
circle of 30.4 m radius, with one string at the center and three strings 
on the circumference. A laser calibration system was developed in order
to determine the precise detector geometry and the 
optical properties of the ice. A brief summary of the
results has recently been published in \cite{2,3}. The results
have shown that the ice at 800 - 1000 m depth is highly turbid. The
 short scattering lengths (between 10 and 20 cm) are consistent with 
being due to a residual population of bubbles of trapped air. 
Deeper in the ice these bubbles 
are expected to be transformed to air hydrate crystals, which have a 
refractive index close to that of ice \cite{uchida} and therefore 
should not cause much scattering. (For a thorough discussion of 
various scattering mechanisms in deep Antarctic ice, see the 
accompanying paper \cite{paper2}.) 

 Surprisingly, the  absorption coefficient
of the ice at wavelengths between 400 and 600 nm was found to
be  smaller by a large factor (greater than 
10 at the shortest wavelengths) than the value for laboratory ice 
\cite{11,17} 
(for a compilation of several sets of experimental data see 
\cite{4}).
The South Pole ice is thus a remarkably clear optical medium,  and
in the 1995-96 season we have deployed 4 new strings to a maximum 
depth of 2000 m, where air bubbles  should  have entirely 
transformed into air hydrates. The analysis of data from this deployment is in 
progress.

Since the reported AMANDA results (especially those of \cite{3}) 
deviate from previously published values by such a large factor, we have felt 
the need to present to the optical physics community the details of 
our experimental method, the proposed explanation
and implications of the long absorption lengths and an analysis of
absorption in dust present in the ice.  Various 
sources of residual scattering at bubble-free depths are discussed in 
\cite{paper2}.

\section{Outline of the method to extract scattering and absorption lengths}

The experimental setup for the measurement of the optical properties 
of the South Pole deep ice is shown in Fig.\,1. Light from a dye laser
(driven by a Nd:YAG laser) was brought down optical fibers to  diffusing
nylon spheres (one per PMT) approximately 30 cm below each PMT.

Pulsed laser light (4 ns FWHM) was fed into the optical fibers at a frequency
of 10 Hz. By using gratings in the dye laser and a monochromator the
wavelength of the light could be varied between 410 and 610 nm, with
a precision of around 2 nm. (At
shorter wavelengths, not enough light could be transmitted through 
the kilometer-long fibers.) After passage down the fiber, the 4 ns pulses
had broadened to around 12 ns. Light was led down to one nylon sphere
at a time, and the response of neighboring  PMTs in the same string
and in the other strings was recorded.

The PMTs (8 inch EMI 9353 tubes with a noise rate in the ice of
about 1.7 kHz)  were operated at a gain of $10^8$, and had
approximately 6 ns resolution, including time smearing in the
coaxial cables that fed the signals up to the surface. The signal 
from the PMT 30 cm from
the flashing nylon sphere  defined t=0 (corrected for run time
through fibers, etc.).

The main purpose of the laser system just described was to serve as
a calibration system for the absolute timings of the PMTs and their relative 
positions in the AMANDA muon detector. Immediately after
the system became operational it was realized, however,  
that the South Pole ice
at depths between 800 and 1000 meters is not an ideal optical medium
for imaging the tracks of muons through their Cherenkov light. Instead
of a narrow arrival pulse (expected, typically, at 87 ns for 20 m distance
between emitting and receiving tube), a very broad structure with maximum
at around 500 ns was observed (see Fig. 2). As very exact timings
are required to recontruct the direction of muons from the sparsely sampled
Cherenkov photons they generate, the observed timing pattern makes
the present array not useful for muon tracking 
(but the array is  used
for other  astrophysical applications \cite{1a}).

However, the presence of the broad time distributions, and our modeling 
of them, have permitted us to make more accurate measurements
of the optical properties of ice in the visible region than have ever been
done in the laboratory. As shown below, the {\em in situ}
laser measurements in large volumes of ice and the random walk nature of the
light propagation permit a clear separation between absorption and scattering,
something which is extremely difficult to study in the laboratory.

In that process, we have independently
discovered a method similar to one recently used in medical and other
applications \cite{patterson}, one difference being the longer time scales 
and distances involved here, another the fact that both the light emitters
(nylon spheres) and receivers (PMT tubes) are immersed in the medium to be
measured. 
The derivation that we will present of the relevant formulas
seems to be new (and simple), and may be of interest for optical physics
researchers. We will also show that a detailed Monte Carlo simulation
of light propagation supports the analytical treatment, and fits the
measured distributions excellently.

\subsection{Random walk description of light propagation}  

The physical picture of photon propagation in the bubbly ice is that of a 
random walk. When a photon leaves
the nylon sphere it passes through clear ice along a straight trajectory 
for some path length until
it hits an air bubble. Then it is refracted (or externally reflected),
travelling in a new direction until it hits the next bubble. 
The wide dispersion of
arrival times at the PMTs must mean that a typical photon has undergone
many scatterings, and therefore a probabilistic analysis of the
random walk should be possible. The problem is simplified by the 
randomness of the medium and the macroscopic
distances of propagation which means that any interference effects can
be ignored.

We define the simple, geometrical scattering length on spherical bubbles as 
\beq
\lambda_{bub}={1\over c_{bub}},
\eeq
with the scattering coefficient $c_{bub}$ given by
\beq
c_{bub}=\int_{0}^\infty {dn_{bub}(r)\over dr}\pi r^2 dr,
\eeq
where $dn_{bub}(r)/dr$ is the number density of bubbles having radius 
between $r$
and $r+dr$;
$\pi r^2$ is
the geometrical cross section. This means that the probability
distribution for the  
 distance $d$ between two successive scatterings 
 is proportional to $e^{-d/\lambda_{bub}}$. We
note that the average geometrical scattering length is
\beq
\langle d\rangle = \lambda_{bub}\label{onefactor}
\eeq
and the second moment is
\beq
\langle d^2\rangle = 2\lambda_{bub}^2\label{twofactor}
\eeq

We only have to incorporate two modifications to the standard random walk
treatments given in the literature (e.g, \cite{chandrasekhar}), namely
absorption of the light beam and the non-isotropy of the scattering 
amplitude
for photons incident on air bubbles. Since the air bubbles themselves
do not cause any absorption of photons, only scattering, it is very
simple to incorporate absorption in the random walk picture. Suppose that a
photon arrives at a PMT after $N$ steps. Then the total path length
is on the average $N\lambda_{bub}$, and absorption can be taken into account
by weighting that particular path by the factor
\beq
e^{{-N\lambda_{bub}\over \lambda_a}}
\eeq
(of course we assume $N>>1$; typically $N$ is on the order
of 1000 in our problem), where $\lambda_a$ is the true absorption length
(i.e. related to true absorption of photons in the medium).

Sometimes it is convenient to discuss absorption in terms of the 
absorption coefficient,
\beq
\alpha_{abs}={1\over \lambda_{abs}}.\label{cabs}
\eeq
This is related to the absorptive (imaginary) part of the index of 
refraction by
\beq
\alpha_{abs}={4\pi {\rm Im}(n)\over \lambda_{L}},
\eeq
where $\lambda_{L}$ is the wavelength of the light source.

Non-isotropy of the scattering from bubbles means that there is a correlation between
successive vectors making up the random walk. If the bubbles are spherical
there is still an azimuthal symmetry, and one can show that (e.g. by 
introducing
rotation matrices which rotate each successive vector to the polar axis)
\beq
\langle \vec r_1\cdot\vec r_{1+k}\rangle =\lambda_{bub}^2(\delta_{0,k}
+\tau^k),\label{taus}
\eeq
where $\tau=\langle \cos\theta\rangle$ is the average of the cosine of 
the scattering angle, and the factor of 2 entering for $k=0$ ($\delta_{0,k}$
is a Kronecker delta function) is 
a reflection of Eq.(\ref{onefactor}) and Eq.(\ref{twofactor}).

In the formula for random walk with absorption introduced as just explained:

\beq
W_N(\vec R)={1\over (2\pi\langle \vec R^2\rangle_N/3)^{3/2}}
e^{-{3\vec R^2\over 2\langle \vec R^2\rangle_N}}
e^{-N\lambda_{bub}\over \lambda_a},
\eeq
it just remains to evaluate:
\beq
\langle \vec R^2\rangle_N =\langle (\sum_{i=1}^N \vec r_i)^2\rangle.
\eeq

By expanding the squared sum in $\langle (\sum_{i=1}^N \vec r_i)^2\rangle$
 one finds
$N$ terms with relative distance $0$ (namely, $|\vec r_1|^2+ 
|\vec r_2|^2+\ldots + |\vec r_N|^2$), $2(N-1)$ terms of relative distance
$1$ ($\vec r_1\cdot \vec r_2 + \vec r_2\cdot \vec r_1 +\ldots +
\vec r_N\cdot\vec r_{N-1}$), and in general $2(N-k)$ terms of relative
distance $k$.

Using then Eq.(\ref{taus}), we find that the factor $2\lambda_{bub}^2$
can be taken outside and we arrive at

\begin{eqnarray} \langle \vec R^2\rangle_N = 2\lambda_{bub}^2\sum_{k=0}^{N-1}(N-k)\tau^k=\\
 2N\lambda_{bub}^2{(1-\tau)+(\tau^{N+1}-\tau)/N\over (1-\tau)^2}.\nonumber
\end{eqnarray}

In the limit of large $N$ we obtain 

\beq
\langle \vec R^2\rangle_N = {2N\lambda_{bub}^2\over 1-\tau}.
\eeq

With the identification $N\lambda_{bub}=c_it$, where $c_i=c/n$ is the
velocity of light in the ice, we obtain the formula corresponding to
the Green's function for the radiative transport of a spherically
symmetric laser pulse emitted at a distance $d=0$
at time $t=0$
\beq
u(d,t)={1\over (4\pi Dt)^{3/2}}e^{-d^2\over 4 Dt}
e^{-c_it\over\lambda_a},\label{u}
\eeq
with $d^2=\vec R^2$, and where the constant of diffusion $D$ is given by
\beq
D={c_i\lambda_{eff}\over 3},\label{diffuse}
\eeq
with the effective scattering length $\lambda_{eff}$ related to 
$\lambda_{bub}$ through the formula
\beq
\lambda_{eff}={\lambda_{bub}\over 1-\tau}.\label{diffuse1}
\eeq

We note that in the expression for $D$ only the refractive scattering part 
of
the optical amplitude needs to be taken into account. If diffraction or
any other small-angle scattering effect is included,
making the geometrical bubble cross section increase from $\sigma$
to $(B+1)\sigma$, then 
\begin{displaymath}
\lambda_{bub}\to\lambda_{bub}/(1+B),\ \tau\to (B+\tau)/(B+1),
\end{displaymath}
 and $D$ is easily seen to be invariant under
these transformations. This has the simple physical 
interpretation that the diffusion power of the very forward-peaked diffraction
has to be negligible compared to the large-angle scattering induced by
refraction on bubble walls. 

The formula (\ref{u}) can be put in a more convenient form as follows using
two dimensionless quantities $y$ and $\kappa$.
Defining $y=t/\sqrt{f}$ with
\beq
f={3\lambda_ad^2\over 4c_i^2\lambda_{eff}}
\eeq
and the dimensionless quantity
\beq
\kappa={\sqrt{3}d\over 2\sqrt{\lambda_a\lambda_{eff}}}
\eeq
one gets
\beq
u(f,y)={C\over y^{3\over 2}}e^{-\kappa(y+1/y)}
\eeq
with
\beq
C={1\over f^{1/4} (4\pi c_i\lambda_{eff}/3)^{3\over 2}}.
\eeq
Using the basic integrals
\beq
\int_0^\infty dy\,y^{-1/2}e^{-\kappa (y+1/y)}=
\int_0^\infty dy\,y^{-3/2}e^{-\kappa (y+1/y)}=\sqrt{{\pi\over \kappa}}
e^{-2\kappa}
\eeq
one finds the average value of arrival time $t$ for a given distance $d$
\beq
\langle t\rangle = \sqrt{f}\langle y\rangle=\sqrt{f}={\sqrt{3}d\over 2c_i}
\sqrt{{\lambda_a\over \lambda_{eff}}},
\eeq
i.e. a linear relation.

In fact, defining the general class of integrals

\beq
f_I(\kappa,n)=\int_0^\infty dy\,y^{-{3\over 2}+n}e^{-\kappa (y+1/y)}\label{kappint}
\eeq
one finds a very convenient recursion relation by taking the derivative
of Eq.(\ref{kappint}) with respect to $\kappa$:
\beq
f_I(\kappa,n+1)=-{d\over d\kappa}f_I(\kappa,n)-f_I(\kappa,n-1),
\eeq
with initial conditions for the recursion given by
\beq
f_I(\kappa,0)=f_I(\kappa,1)=\sqrt{\pi\over\kappa}e^{-2\kappa}
\eeq
(the equality of $f_I(\kappa,0)$ and $f_I(\kappa,1)$ , or more
generally, $f_I(\kappa,n)$ and $f_I(\kappa,1-n)$
is easily demonstrated by letting $y\to 1/y$ in the integral.)

 Using
the arrival time distribution expressed in this scaling form, one can
easily calculate all moments of the distribution, e.g., the
expectation value \beq
E(y)=1,
\eeq
and the variance
\beq
V(y)=E(y^2)-E(y)^2={1\over 2\kappa}.
\eeq
The scaling form of the distribution was used in the numerical
fitting procedure for the arrival times of the laser pulses.  

The value of time $t_{peak}$ at
which the distribution peaks is given by
\beq
t_{peak}={\sqrt{3}d\over 2c_i}\sqrt{{\lambda_a\over\lambda_{eff}}}
\left(\sqrt{1+{3\lambda_a\lambda_{eff}\over 4d^2}}-{\sqrt{3\lambda_a
\lambda_{eff}}\over 2d}\right).
\eeq
For the case at hand, with
$\lambda_a >> \lambda_{eff}$, this also gives an essentially linear
dependence of $t_{peak}$ on $d$
 (compared to $t_{peak}\propto d^2$ for a random walk 
process without attenuation). 

\subsection{Application to muons traversing bubbly ice}

The results for the Green's function derived here can also be used to obtain an expression
for the time distribution of Cherenkov photons coming from a straight muon track, 
for which the closest approach (``impact parameter'')
$d_i$ to the PMT at $t=0$ 
is much larger than the
scattering length $\lambda_{eff}$ (i.e. in our case greater than a meter
or so).
Introducing $a=4\lambda_{eff}c_i/3$ and a cut-off length  $l$ (the ``size''
of the detector) one finds
\beq
I(d_i,t;l)=N\sqrt{{c_3\over c_1}}\int_{max\left(0,\sqrt{c_3/c_1}
(t-l/c_0)\right)}^{\sqrt{c_3/c_1}(t+l/c_0)}{dz\over 
z^{3/2}}e^{-\left(\sqrt{c_1c_3}(z+1/z)+c_2\right)},
\eeq
where $N$ is a normalization factor, $c_0$ is the velocity of the muon 
(close to the velocity of light in vacuum), and
\beq
c_1=(d_i^2+c_0^2t^2)/a,
\eeq
\beq
c_2=-2c_0^2t/a,
\eeq
\beq
c_3=c_0^2/a+c_i/\lambda_a.
\eeq

The total flux $F_\mu(d_i)$ received at a PMT of area $A_{PMT}$ 
from a muon 
generating
$N_C$ photons per unit track length passing the PMT at impact parameter $d$
can be obtained (for $d_i>> \lambda_{bub}$)  by integrating this
expression over time and taking the limit $l\to \infty$. We find

\beq
F_\mu(d_i)={3N_CA_{PMT}\over 8\pi\lambda_{eff}}K_0\left(d_i\sqrt{
{3\over \lambda_{eff}\lambda_a}}\right),
\eeq
where $K_0$ is a modified Bessel function of the second kind.

In applications where directional information and time structure at 
the sub-microsecond level are not important, such as the detection
of neutrinos from supernovae where relevant time scales are on
the order of seconds, the present AMANDA detector in bubbly
ice is perfectly functional. A neutrino-induced event from a SN
explosion typically generates a positron track of only 10 cm, and
can on the diffusion scale $\sqrt{\lambda_a\lambda_{eff}}$ be regarded
as pointlike. Then the total response function (flux per unit area
integrated over time) for $N_{SN}$ photons (typically of the order
of 3000 per positron) injected at distance $d$ is given by

\beq
I_{SN}(d)={3N_{SN}\over16\pi\lambda_{eff}d}
e^{-\sqrt{3\over\lambda_a\lambda_{eff}}d}.
\eeq

We can define an effective volume of ice seen by a PMT by integrating
$A_{PMT}$ times this flux over all space (the error caused by using
the formula at small $d$ where it is not strictly valid can be shown
to be small, as it contributes a very small fraction of the result).

We then obtain

\beq
V_{SN}={N_{SN}A_{PMT}\lambda_a\over 4}.
\eeq 

This result could (apart from the numerical factor) be anticipated
on dimensional grounds using the observation that scatterings do not change
the average photon flux (thus the result should not depend on the scattering
length). For a real detector, correction factors related to the quantum
efficiency of the PMTs and Poisson statistics of detected photons have to 
be added. Since the Cherenkov flux depends strongly on wavelength ($\sim
1/\lambda^2$) as do $\lambda_a$, the quantum efficiency and to some
extent $\lambda_{eff}$, an integration over wavelength has to be 
performed in addition. 

\subsection{Monte Carlo simulation of light propagation in bubbly ice}

To verify that the general picture and derivations above are correct,
we have also performed extensive Monte Carlo simulations of light
propagation through the ice.  
Thanks to the 
existence of widely separated length scales, $d\sim \lambda_a>> 
\lambda_{eff}>> \lambda_L$, where  $\lambda_L$ is the 
wavelength of the 
laser light, a fairly simple optical treatment is adequate to obtain an 
estimate of  $\lambda_a$ and $\lambda_{eff}$  good to a few percent
accuracy or better. In fact, the estimate of $\lambda_a$ is given quite 
directly by the fall-off of the arrival times of photons for large times, 
and is very insensitive to the uncertainties in the determination of 
$\lambda_{bub}$. To determine $\lambda_{bub}$ one 
 needs the average value of the
cosine of the scattering angle as input. 
One cause of uncertainty is  that the
bubbles may not be smooth enough to be regarded as perfect spheres. In that 
case, the diffusing ability could be greater (i.e. $\langle \cos\theta\rangle$
smaller) and $\lambda_{bub}$ consequently larger. Thus, we are able to
determine $\lambda_{eff}$ quite accurately from laser data, but the relation 
between
$\lambda_{eff}$ and the geometrical $\lambda_{bub}$ contains uncertainties.
A direct measurement of the scattering function of air bubbles contained
in the ice cores taken from the Antarctica would obviously be of importance
for this problem.

Since photons from the coherent laser source scatter on the average 
between 500 and 1500 times on bubbles before arriving at the PM tubes, the 
initial state in any single scattering can be regarded as being completely 
incoherent in phase, direction and polarization. We have 
thus employed simple geometrical optics for the scattering process 
incorporating singly refracted, externally reflected and totally reflected 
light with the relative strength of the components given by the Fresnel 
coefficients. (In fact, we have also performed full Mie scattering
calculations and verified that the value of $\tau$  obtained from
this simple analysis is correct.)
 We take into account the change of index of refraction of 
the air in the bubble due to the pressure and temperature. 
For the ice we use the 
wavelength-dependent values of 
the refraction index given in \cite{9}. The bubbles are assumed to be
spherically symmetric,  in agreement with observations at
these depths. 

 Although a more complete treatment 
of the scattering could be implemented in the Monte Carlo, we emphasize 
that this, according to Eq.\,(\ref{diffuse1}), only amounts to a 
renormalization of the relation between the effective scattering length
$\lambda_{eff}$ and the geometrical bubble scattering length
$\lambda_{bub} $ (with the latter defined as an average over sizes and
shapes of air inclusions), which is not 
important for
discussion in this paper. (It is of importance, though, when discussing
the implications of our results, e.g., for glaciology.)
With our approximations we find for the 
quantity $\tau$  entering 
 Eq.\,(\ref{diffuse1}), a value around 0.75. This is slightly 
 wavelength-dependent as the refractive index of ice increases 
 by some 2\ \% when
 going from 600 to 300 nm. The Cherenkov angle $\arccos(1/(n_i\beta))$
 (where $\beta$ is the muon velocity in units of the speed of light in vacuum)
 will also increase by some 4\ \% in going from the longest to the 
 shortest wavelengths. (Since the intensity spectrum of the Cherenkov radiation
 varies as $1/\lambda^2$,  the shorter wavelengths are  most
 important for muon detection. Below 300 nm, however,
 the absorption in ice increases rapidly and the detection efficiency of the
 optical modules drops.)

\subsection{Fits to the experimental distributions}

In making global fits to the laser calibration data we used special
routines that computed maximum likelihood fits to the observed
distributions. The chi-squared method was used to give starting values
for the maximum likelihood fit and to give a figure of merit of
the fit. The Davidon variance metric algorithm \cite{Davidson} was 
used for the minimization and the statistical errors were extracted 
from the variance matrix of the maximum likelihood procedure.
The function used was
formulated as
\beq
g(t)={x_n\cdot E_n\cdot\Delta x\cdot e^{-T}\over y^{3\over 2}{\cal N}},
\eeq
where $x_n$ are the normalization constants for each measured arrival
time distribution ($x_n\approx 1.0$), $E_n$ the number of events in
distribution $n$, $\Delta x$ the bin size used in all distributions,
$$T=\kappa\biggl(y+{1\over y}\biggr),$$
$$y={t\over \sqrt{f}}\ {\rm or}\ {t-\Delta t_m\over \sqrt{f}},$$
where $\Delta t_m$ allows for a shift in the definition of time zero
for each emitting tube (this this depends on details of the PMT such as
the value of the high voltage etc). We
allowed for uncertainties in the vertical positions of the strings as that
was difficult to determine during deployment.
The initial chi-squared sum we minimized was
\beq
\chi^2=\sum_{n=1}^{n_{dis}}\sum_{i=1}^{n_{bins}}{(y_{i,n}-g(t))^2\over y_{i,n}},
\eeq
where the $i$th bin in the $n$th distribution has a content of $y_{i,n}$
events and the theoretical prediction for this bin is $g(t)$.

There is a potential problem with this  laser calibration method 
due to the fact that it is
in general impossible to resolve several individual photons from the same
laser pulse hitting the same PMT. Since the arrival time recorded by
the receiving PMT is set by the first
photon, one may tend to suppress long arrival times, which could underestimate
the absorption length.
To overcome this problem,  we monitored 
the total number of emitted laser pulses $N$. For
each receiving tube we then estimated the average number of photons
reaching this detector from the zero class of a Poisson distribution.
If the distribution has $n$ entries the Poisson parameter $\mu$ is
given by
\beq
P(0)=e^{-\mu}={N-n\over N},
\eeq
i.e.
\beq
\mu=\ln \biggl({N\over N-n}\biggr).
\eeq
The variable used to select non-saturated data is the sum of the zero class
and the class with one photon,
\beq
\alpha=P(0)+P(1)=e^{-\mu}+\mu e^{-\mu}=e^{-\mu}(1+\mu),
\eeq
i.e. $1-\alpha$ is the fraction of events
expected to have two or more photons reaching the detector and thus
causing saturation. To avoid the saturation problem, we demanded that
$1-\alpha$ was less than 1 \%. To get a statistically significant 
estimate of the optical parameters only time distributions where at least 
2000
hits were registered by the receiving PM were selected. This left 
almost 500 time distributions for various wavelength and distance 
combinations.

In Fig.\,2 we show the
distribution of arrival times as measured and  as fitted to  
the analytical expression (Eq.\,~13) for our random walk model,
 for an emitter and a receiver 
separated by 10 m. The results for wavelengths 410 and 610 nm are
shown as open and filled circles, respectively. The fits shown 
give the same value for the scattering length $\lambda_{bub}$ (around
10 cm) for the two cases, but absorption lengths of around 230 m and 
10 m, respectively.
The quality of the analytical fit to the data
is quite remarkable. We thus conclude 
that the physics behind the laser calibration data is well understood, 
and that we have, in fact,  at our disposal a very powerful tool to make
{\em in\ situ} measurements of 
important properties of the ice such as $\lambda_a$ and $\lambda_{bub}$.
In particular, $\lambda_a$ and $\lambda_{eff}$ are essentially uniquely 
determined by the temporal distribution of arrived photons at each 
particular PM tube, once the geometry of the detector array is fixed.

One may notice that the values of $\lambda_a$ and $\lambda_{eff}$ obtained
through this analysis are essentially spatial averages over the diffusion scale
$\sqrt{\lambda_a\lambda_{eff}}$. Since we let these parameters depend on
depth in the analysis we are using an ``adiabatic approximation'' that
should be excellent if the quantities vary slowly with depth. The goodness
of the fit to the function (\ref{u}) for all wavelengths and all depths
shows the self-consistency of this scheme.

\section{Summary of Experimental Results}

In this Section, we briefly summarize the results on absorption length as
a function of wavelength obtained by the AMANDA calibration system at
two depths, around 830 and 970 m depth, respectively. In \cite{3} 
results are also given for an intermediate region around 910 m, with
results similar to those at the shallowest depth. The results
are shown in Table 1 (for details, see \cite{3}). The errors given
are only statistical; a systematic error estimated to be around 5 \%\ 
(dominated mainly by the uncertainty of the exact geometry of the
AMANDA array should be added.

\section {Modeling the Op\-tical Pro\-perties of High\-ly Pure South Pole Ice}

	Using the results presented in Section 3, we next analyze the intrinsic optical 
properties of the South Pole ice and estimate the absorption due to solid impurities present 
in the ice. In the discussion we compare our results for South Pole ice with measurements 
made on less pure ice \cite{4,11,17}, on extremely pure water \cite{12}, 
and on LiF \cite{18}  as an
example of a highly transparent solid.

\subsection{Intrinsic absorption in highly pure ice}

	Our discussion is based on features shown in Figs. 3 to 5. 
	We focus attention on 
three regions of the approximately V-shaped spectra with their
 minima in the visible region. 
On the ultraviolet side the absorption in both ice \cite{4,11,17} 
and ionic crystals \cite{18}
 decreases 
exponentially at wavelengths somewhat longer than that corresponding 
to the electronic band gap 
energy ($\sim 8$ eV for ice, $\sim 13$ eV for LiF). The slope of
 this so-called Urbach tail is 
believed to be governed by exciton-phonon interactions; for LiF 
 it becomes more 
shallow with increasing temperature \cite{18}.

	On the long wavelength side of the minimum, absorption rises
	 approximately 
exponentially as a function of wavelength, but the window of transparency 
is much 
narrower for ice than for ionic crystals. Ionic crystals are transparent
throughout the entire 
visible region, and their exponential absorption spectrum in the infrared 
is the result of 
multiphonon processes, in which a photon is absorbed with the emission 
of $n$ phonons. 
Phenomenological models \cite{20} lead naturally to an exponential dependence
 of the absorption 
coefficient on wavelength, in agreement with measurement over six 
orders of magnitude, 
for $n$ up to at least seven phonons. The argument in the case of ionic 
crystals is that the contributions of
overtones are generated by processes that depend on some coupling constant, 
with the $n$th 
overtone going as (coupling constant)$^{n-1}$. This leads naturally to an 
exponential dependence 
on overtone number.

In contrast, ice is a molecular crystal in which the $H_2O$ molecules are
connected by weak $OH-O$ hydrogen bonds. Each $O$ atom in the lattice
is coordinated to four $H$ atoms, forming strong intramolecular
bonds with two of them and hydrogen bonds with the other two. Ice
 absorbs much more strongly than do ionic crystals in the red region of 
 the visible. The  
reason is that ice has additional degrees of freedom associated with the
 stretching and 
bending of H-O-H bonds in individual water molecules. These internal 
modes lead to 
absorption in the red region. The fundamental stretching mode peaks 
at $\sim3$ $\mu$m. Numerous 
overtones and combinations of stretching and bending modes with each 
other and with the 
lattice modes lead to a roughly exponential slope from a minimum at 
about 400 nm to a 
large peak at 3 $\mu$m. The extension of this exponential to values of
the absorption coefficient 
below $10^{-2}$ m$^{-1}$ can be seen only if the ice is very pure 
and if the
 measurement technique is 
able to discriminate absorption against scattering.
 
	By application of the same reasoning that applies to multiphonon 
	absorption, we 
suggest that intramolecular vibrations depend on a coupling constant 
and that the overtones 
decrease as increasing powers of the coupling constant, thus leading 
to an exponential 
dependence of absorption on wavelength in the red and infrared. Since
 this conjecture 
assumes no growth in the number of processes with overtone number, the 
actual rate of 
increase of absorption with wavelength may, for large overtone numbers, be 
less rapid than exponential. (We will make use of this assumption of 
an exponential
absorption in Section 4.5 when we estimate the contribution of dust
to absorption.)

	Impurities, dislocations, vacancies, and their associated electronic 
	defects contribute 
to absorption and scattering. Almost no measurement techniques, with the 
exception of our 
pulsed laser technique, are designed to be able  to distinguish 
scattering from absorption. 
These defects 
tend to fill in the minimum in the spectrum to a degree that depends on their 
concentration and size 
distribution.

	Figure 3 compares absorption data and the calculated magnitude of
	 Rayleigh 
scattering for ice (solid curves and open symbols) and water (dashed 
curves and solid points) at wavelengths 
near the absorption minimum. The most striking feature of the data is 
the large discrepancy 
between values of the absorption near the minimum. For water, the 
absorption decreases in 
order from the water of Lake Baikal \cite{21}, to the Pacific 
Ocean \cite{22} (the DUMAND point), to 
``pure sea water'' (with no particulate content) 
as summarized by Smith and Baker \cite{23}, to the 
curve labeled ``purest water'', 
which gives the results of Quickenden and Irvin \cite{12}, who took 
extraordinary measures to 
eliminate both inorganic and organic impurities, and finally to the point 
labeled SNO. The SNO result is a recent 
measurement by No\"el and Mes \cite{36}, who went to great lengths to
 purify the water to be used 
in the Sudbury Neutrino Observatory experiment. 

The Lake Baikal group has recently used a pulsed laser at 475 nm to measure
scattering separately from absorption. They find values of 
$\lambda_{scat}\approx 8-23$ m$^{-1}$, with $\langle\cos\theta\rangle\sim
17^o$ \cite{baikalscatt}.
None of the authors of the other studies on 
water made a case for having been able to distinguish Rayleigh 
scattering from absorption. 
In fact, the data for purest water approach asymptotically (to 
within a factor two) the 
calculated Rayleigh scattering line for pure water.

	The measurements in Fig.\,3 labeled ``lab ice'' represent the 
	contributions of a 
number of researchers \cite{11,17},  none 
of whom took steps to measure scattering. In contrast, the AMANDA 
experiment measured both the scattered and absorbed components 
separately for ice that is 
probably the purest naturally-occurring solid material on earth. 
(The concentration of 
impurities in South Pole ice is discussed in Ref.\, \cite{24}.) At 
the minimum in the curve for lab 
ice, the absorption measured by AMANDA is about an order of magnitude 
lower than the 
Grenfell and Perovich \cite{11}  data. Fig.\,4 shows the region in more detail, with
$\lambda^{-1}$
 as the abscissa  on a linear scale so that the 
exponential dependence can be more clearly noted.

	The reader may have noticed that, in the interval $\sim
	0.2$ to $\sim 0.45\ \mu$m, the data for lab 
ice in Fig.\,3 have a slope approximating the Rayleigh $\lambda^{-4}$ law,
 but about two orders of magnitude 
higher than the calculated contribution of Rayleigh (or thermal) 
scattering for pure ice 
indicated by the solid straight line. In Section 4.2 and in \cite{paper2}
we suggest 
that the apparent absorption of 
lab ice in this interval may be due to Rayleigh scattering from 
defects in lab ice.
	The dotted line in Figs. 3 and 4 is an exponential fit to 
	the  data of Grenfell and Perovich \cite{11}
in the interval 
0.6 $\mu$m to 0.8 $\mu$m. It represents our conjecture as to 
the intrinsic absorption
 of ice in the red 
region in the absence of impurities.
 
	Fig.\,5 compares the absorption in ice and a number of 
	transparent solids over a broad range of 
wavelengths. We draw attention to the lines representing
 fourth-power laws, $\lambda^{-4}$,
 which 
suggest scattering from crystalline defects with dimensions 
much smaller than the 
wavelength.  The techniques used to obtain these data 
could not have distinguished scattering from absorption. Thus, 
it is tempting to conclude 
that the portions of the absorption curves that exhibit the $\lambda^{-4}$
 dependence are mainly due to 
impurity scattering instead of being due to absorption. In \cite{paper2} we  
estimate the 
concentration, size, and type
 of defects necessary to account for the $\lambda^{-4}$
 fits to the data.

\subsection{Absorption by solid dust particles in ice}

	We now focus on the AMANDA absorption data\,\cite{3} taken at depths
	 of 830 m (shown 
in Figs.\,3 and 4 as triangular symbols) and at 970 m (shown in
 Figs.\,3 and 4 as square symbols). 
Additional data, taken at a depth of 910 m, were indistinguishable 
from those taken at 830 
m. The data\,\cite{3} show convincingly that the absorption deviates
 more from the exponential 
extrapolation at 970 m than at 830 and 910 m. Our interpretation is 
that, in addition to the 
intrinsic absorption by perfect ice, dust particles with a greater 
concentration at 970 than at 
830 and 910 m absorb light with a coefficient that decreases with 
increasing wavelength. This 
interpretation is strongly supported by our model of age vs depth 
of South Pole ice, by 
means of which we predicted dust concentration to increase to a 
maximum at a depth 
slightly greater than 1000 m (ref.\,\cite{13}). 

We now derive the 
absorption coefficient as a 
function of wavelength for dust in South Pole ice, and we determine 
the ratio of dust 
concentration at 830 m to that at 970 m.
	When trying to extract absolute values for the contribution of 
	solid impurities to the 
absorption coefficient, one encounters the problem of estimating the 
uncertainty involved in 
the exponential extrapolation of the intrinsic absorption coefficient 
of ice shown as the 
dotted line in Figs.\,3 and 4. We developed the following way to 
determine the wavelength 
dependence of the absorption of solid impurities, normalized to 
some particular 
wavelength, independent of the unknown intrinsic ice absorption. We 
write the measured 
absorption coefficient $\alpha_{expt}(h,\lambda)$ as
\beq
\alpha_{expt}(h,\lambda)=\alpha_{int}(\lambda)+\alpha_{dust}(h,\lambda)=
\alpha_{int}(\lambda)+c_{dust}(h)
\sigma_{dust}(\lambda),
\label{c_exp}
\eeq
which exploits the facts that the intrinsic ice absorption coefficient 
$\alpha_{int}(\lambda)$
 and the 
absorption cross section of solid impurities $\sigma_{dust}(\lambda)$ 
are independent of depth, and the 
concentration of impurities, $c_{dust}(h)$, is independent of wavelength. This 
assumes that the nature of solid impurities is independent of depth and 
only the 
concentration changes\footnote{Royer et al.\,\cite{14} inferred from the optical 
scattering properties of 
melted ice core samples from Dome C that the particle size distribution 
and complex 
refractive index of dust show no depth dependence, although it is not
clear to us that these conclusions were very strongly supported by their 
data.}.

	By using eq. (\ref{c_exp}) for two different wavelengths and two 
	different depths, we can 
eliminate the unknowns $\alpha_{int}$ and $c_{dust}(h)$ and express 
the ratio of the absorption 
cross section
 at two wavelengths in terms of only experimentally measured quantities. The 
result is
\beq
{\sigma_{dust}(\lambda_1)\over \sigma_{dust}(\lambda_2)}=
{\alpha_{expt}(h_1,\lambda_1)-
\alpha_{expt}(h_2,\lambda_1)\over \alpha_{expt}(h_1,\lambda_2) - 
\alpha_{expt}(h_2,\lambda_2)}.
\eeq 
We will use for the analysis only the data for the six wavelengths 
that give the smallest
statistical error. 
For $h_1$ we use data from phototubes at 830 m (around AMANDA level 
3) and for $h_2$ we use data at 970 m (around AMANDA level 17). 
We normalize all impurity absorption cross sections to
$\sigma_{dust}$(475 nm), which has the most 
statistics of all the distributions. 
Figure 6 shows the result of this analysis. To test the natural 
hypothesis that the absorbing impurity is atmospheric dust 
that accumulated in the ice by 
precipitation onto the surface at the time the ice formed from 
snow\,\cite{14}, we have compared 
our results with  measurements of the absorption spectra
of various samples of atmospheric dust  at  
U. S. locations\,\cite{15}. As can be seen, our results are consistent 
with  the hypothesis 
that we do observe absorption by dust. However, it must be noted that
the model-independent analysis
does not have small enough error bars to exclude, e.g., a constant 
absorption
coefficient.

	In Fig.\,6 we have also
included estimates based on  the assumption that the intrinsic 
absorption spectrum of dust-free ice follows the exponential shown by a 
dotted line in 
Fig. 4. The good fit shown in that Figure was obtained by the expression
\beq
\alpha_{int}(\lambda)=81\exp(-\lambda_0/\lambda)\ \ {\rm cm}^{-1},
\label{expslope}
\eeq
with the value $\lambda_0\sim 6.7\ \mu$m being determined not only by a 
fit to the measured absorption in the near $IR$, but also requiring 
self-consistency
of the {\em Ansatz} (\ref{c_exp}). That is, the values obtained for 
dust concentration
 $c_{dust}(h)$ using the experimental values $\alpha_{expt}(h,\lambda)$ 
 and the fitted $\alpha_{int}(\lambda)$ should not depend on $\lambda$; 
 neither
 should the derived $\sigma_{dust}(\lambda)$ depend on $h$. 
 
 Using (\ref{c_exp}), we can derive the following expression for the 
 ratio of the dust concentration at the lower level of the detector to 
 the upper level:
 \beq
 {c_{dust}(970\ m)\over c_{dust}(830\ m)}=
{\alpha_{expt}(970,\lambda_1)-
\alpha_{int}(\lambda_1)\over \alpha_{expt}(830,\lambda_1) - 
\alpha_{int}(\lambda_1)}.
\eeq
We see in Fig.\,7 that 
all wavelengths give the consistent result $\sim 1.4$ for this ratio.
 
Subtracting now the fitted intrinsic absorption coefficient from the 
measured one at each wavelength, we 
 finally obtain values for the absorption length (i.e. the inverse of 
the absorption coefficient) due to dust at the two AMANDA levels. 
We find that the absorption length due to dust increases with wavelength and 
decreases with depth, with typical values being 200-300 
meters, as shown in Fig.\,8. This is consistent with estimates based on measured  dust content and 
composition in South Pole ice. It must be kept in mind, though, that 
the exponential fit to the intrinsic absorption coefficient is only a
phenomenological procedure. To the statistical error bars in Fig.\,8 must
therefore be added a systematic uncertainty of unknown magnitude.

We note that our results at 830 m depth give values for the absorption
coefficient that represent the purest ice ever measured. Since a 
residual dust component is known to exist also at these depths, the 
results for $\alpha_{abs}$ must be interpreted as upper bounds for the 
intrinsic absorption coefficient of pure ice.

For completeness, we give in Appendix A a brief description of scattering 
on dust grains,
expected to be the dominant contribution to the turbidity at larger, 
bubble-free depths.

\subsection{A three-component model of absorption in dusty ice}

To summarize the results of this section, Fig.\,9 shows a three-component
model which we expect to apply to absorption by ice with an extremely low
content of dissolved impurities. Component 1, which dominates in the 
ultraviolet, is the exponential Urbach tail, given by
\beq
\alpha_{UV}=A_{UV}\exp(-0.4818\cdot\lambda),
\eeq

where $A_{UV}$ is proportional to the density of insoluble dust in ice. 
As with ionic solids, the slope of the Urbach tail may show a weak dependence
on measurement temperature.

Component 2, which dominates in the near-UV and blue, is the contribution
of insoluble dust, which we assume to have a size distribution similar to
that of the aerosols studied by Lindberg and co-workers \cite{15}.
Fitting their absorption data to a power law in the region $300\leq\lambda
\leq 700$ nm leads to a dependence proportional to $\lambda^{-2}$ for
measurements made both in 1974 and 1994. We assume the same $\lambda^{-2}$
dependence for dust in South Pole ice:
\beq
\alpha_{dust}=A_{dust}\lambda^{-2}.
\eeq

Component 3 is the exponential rise in the red and infrared, as parametrized
in Eq.\,(\ref{expslope}).

From section 4.2, we take $A_{dust}$ at a depth of 970 m to be 1.4 times that
at 830m. The family of curves in Fig.\,~9 is for values of $A_{dust}$
increasing incrementally by factors of 1.4.

None of the curves in Fig.\,~9 fits the data for laboratory ice, nor 
should they. as we noted earlier, the contribution of scattering was not
distinguished from absorption in those experiments \cite{4,11,17}. Even
if scattering had been excluded, the laboratory ice had been made from
not very pure water and thus probably contained dissolved impurities
which contributed their own absorption bands to the ice.

\section{Discussion and Conclusions}

 The pulsed laser method, applied via optical fiber to 
deep ice, provides a 
highly successful method of independently measuring absorption 
and scattering as a 
function of wavelength, over the interval 400 to 650 nm. 

	 Previous laboratory studies of absorption of ice at visible 
	and near-visible 
wavelengths where the scattering length is comparable to or less than 
the absorption length 
are probably in error due to failure to determine the two parameters 
separately and probably also to a greater concentration of impurities
in ice made from laboratory water than in Antarctic ice. Our 
work has shown that the intrinsic absorption of pure ice decreases 
to values less than a few times
10$^{-3}$ m$^{-1}$ at wavelengths between 300 and 400 nm. Our 
pulsed laser results support the conjecture that the absorption 
increases exponentially 
with wavelength in the red region. We think it would be worthwhile 
for quantum 
chemists to seek a theoretical basis for the exponential behavior.

The optical measurements by the AMANDA collaboration, initially 
intended for calibration purposes, have revealed surprising 
properties of pure ice. The remarkably low intrinsic absorption
in the visible and near UV region may lead to  interesting applications
for physics and astrophysics.
Our simple three-component model for absorption by pure but dust-bearing ice
(Fig.\,~9) will be subjected to a critical test when the data at depths
1500 to 1900 m taken in 1996 become available. In those experiments
we will measure absorption and scattering at wavelengths of 337, 350, and 380
nm as well as in the region 410 to 650 nm.

\vskip .3cm
{\bf \large Acknowledgments} 
\vskip .3cm
We are grateful to Y. He and A. Westphal for useful comments on the 
manuscript.
L.B.  acknowledges support from the Swedish Natural Science Research Council 
and
the Stockholm University - UC Berkeley exchange program. 
P.B.P. acknowledges support under the National Science Foundation grant
PHY-9307420 and the Uppsala University - UC Berkeley exchange 
program. The AMANDA collaboration is indebted to the Polar Ice Coring
Office and to Bruce Koci for the successful drilling operations, and 
to the National Science Foundation (USA), the Swedish
Natural Science Research Council, the K.A. Wallenberg Foundation, the 
Swedish Polar Research Secretariat, and the German Electron 
Synchrotron Institute DESY for their support. 
\newpage
\begin{center}{\Large\bf Appendix A. Simulations of scattering on dust grains\\
in Antarctic ice}
\end{center}
\vskip .2cm
Although we have reasons to believe that the optical properties of 
Antarctic ice will be dramatically better at depths where the air 
bubbles  have  transformed into clathrates, for future use we want 
to estimate the 
residual scattering on ice crystal boundaries, clathrates, 
dust and other impurities, 
as well as
Rayleigh scattering in the ice itself.

It is only dust (including soot) of these sources that contributes
measurably to absorption. We start by briefly discussing 
the scattering properties
of dust, given the absorption coefficients derived in the previous Section.

The total concentration of insoluble impurities in recently deposited 
South Pole ice has been estimated
to be around 15 ng/g \cite{Kumai}. The size distribution of 
dust grains 
has also been measured \cite{gayley,mosley} at the South Pole. 
At another Antarctic
location (Dome C), deep samples have been taken and analyzed using  a
scanning 
electron microscope (giving the number {\em vs} radius distribution), 
a 
Coulter counter (giving the volume distribution), and laser 
nephelometry (giving
the combined scattering function of a melted ice sample) \cite{14}. 
In the latter 
work, the data could be well fitted by a log normal distribution with a 
modal radius of $0.25\ \mu$m (for the number to log radius size 
distribution). Using this size distribution and applying Mie 
scattering theory,
 they could get satisfactory agreement with the 
measured scattering function, except at very large angles, where perhaps 
effects of the non-spherical shapes of dust grains appear.

We have developed Mie scattering programs to verify the findings of Royer 
et al.\,for the scattering function. However, when simulations for a 
large detector like AMANDA have to be done it is useful to 
find approximations 
that give results much faster than CPU-time consuming Mie calculations on an 
event-by-event basis.

We have found that a surprisingly good approximation to the results 
in \cite{14}  
can be obtained by simply using first-order geometric optics 
(neglecting internal
reflections) on dust grains, i.e., by just calculating the deviation 
angle according
to Snell's law. This is only a function of where on the beam-facing surface 
of the sphere
the light ray enters. We generate a 
radial variable $r$ at random by setting
$r=\sqrt{x}$ ($x$ is a random number distributed between 0 and 1 
and the square root makes the geometric weight for a given $r$ proportional 
to $r$).
The scattering angle for the Monte Carlo simulation can simply be taken as
\beq
\delta\theta=2(\arcsin(r)-\arcsin(r/n_{rel})),\label{simple}
\eeq
where $n_{rel}$ is the relative refraction index ($\sim 1.18$) between 
dust and ice.

The expression (\ref{simple}) has the drawback that it underestimates 
  large-angle scattering somewhat. In contrast to the
multiple scattering case on air bubbles discussed before, where only
$\tau=\langle \cos\theta\rangle$ is of importance, for reasonable 
possible spacings between PMTs in a working detector only a few scatterings
on dust will occur and therefore the full scattering function is
more important. We  have thus chosen a strategy 
which is based on the Henyey-Greenstein approximation 
to the scattering function:
\beq
{d\sigma\over d(\cos\theta)}={1-\tau^2\over 
(1+\tau^2-2\tau\cos\theta)^{3\over 2}}.\label{henyey}
\eeq
This very accurately reproduces the true scattering function once the 
value of $\tau=\langle \cos\theta\rangle$ is known. To find the latter, we 
calculated it using Mie theory, with the known properties
of the dust in South Pole ice as input. For the values of Dome C \cite{14},
we find $\tau \approx 0.92$. For use in a simulation program, 
Eq.\,(\ref{henyey})
must be integrated and inverted. Setting  now $r=x$ ($x$: random number 
uniformly distributed between 0 and 1), we find that a distribution 
of the
form 
is obtained by choosing
\beq
\cos\theta={2r(1-\tau+ r\tau)(1+\tau)^2\over (1-\tau+2 r\tau)^2}-1.
\eeq

The $1+\cos^2\theta$ distribution applicable to Rayleigh scattering is 
similarly
generated by choosing
\beq
\cos\theta=2(\kappa^{1\over 3}-{1\over 4}\kappa^{-{1\over 3}}+{1\over 2})-1,
\eeq
where
\beq
\kappa=-{1\over 4}+{1\over 2}r +{1\over 8}\sqrt{5-16r+16r^2}.
\eeq

There is a difference between  the measured dust size distributions at Dome C 
and at the South Pole. In the latter case, the modal radius is around 
$0.1\ \mu$m (compared to $0.25$ at Dome C). This makes the scattering 
distribution
markedly wider as a consequence of the approach to
the Rayleigh scattering regime, which is isotropic on the average.

It is 
customary to define the albedo $\omega$ as the ratio of scattering to 
extinction (i.e., scattering plus absorption) cross sections. The 
absorption in dust depends on the imaginary part of the refractive index, 
which has been measured to be typically  between $-0.003$ and $-0.007$ for
atmospheric dust \cite{imdust}. From the Mie 
computations (using again a log-normal size distribution)
  we obtain  values of the albedo
between  $0.9$ and $0.96$ both at 500 and 350 nm, when the real part of the 
refractive index of dust varies between $1.53$ and $1.56$ and the imaginary
part between $-0.003$ and $-0.007$. This is not too different from the
values inferred from the Dome C data in ref. \cite{14}. 
The values of $\tau =\langle\cos\theta\rangle$
are, however, much smaller: $\sim 0.70$ for 500 nm and $\sim 0.74$ for 
350 nm (not very much dependent on the value of the imaginary index). 
The results
using a so-called Junge law distribution $dn/d(log r)\propto r^c$, 
with $c\sim -1.9$
are similar, except that the albedo depends more strongly on the 
imaginary part
of the refractive index. This seems to be due to the effects of 
Mie-type resonances
whose exact location in size parameter depends on both the real 
and imaginary parts
 of the index. The values of the effective scattering power 
 (defined analogously
to Eq.\,(\ref{diffuse1})) are relatively stable and correspond to 
$\omega$ and
$\tau$ both being around 0.8 - 0.9.

We note that typical scattering lengths on dust grains in the 800 - 
1000 m depth range are expected to be of the order of 50 m. Therefore,
scattering on air bubbles dominates completely. For greater depths, 
however, the results of this Appendix should be of use.

\newpage
\null
\vskip .3cm
{\bf\Large Figure Captions}
\vskip .3cm

1. Method for measuring scattering and absorption in deep ice: pulses at a desired 
wavelength are sent down one of 80 optical fibers to a diffusing sphere 
mounted immediately underneath 
a photomultiplier (PMT) module,
and the distribution of arrival times is measured at one of the neighbouring
PMTs.

2.  Distribution of arrival times as measured and  as fitted to  
the analytical formula Eq.\,~13 for an emitter and a receiver 
separated by 10 m. The results for wavelengths 410 and 610 nm are
shown as open and filled circles, respectively. The fits shown 
give the same value for the scattering length $\lambda_{bub}$ (around
10 cm) for the two cases, but absorption lengths of around 230 m and 
10 m, respectively.

3. Comparison of absorption spectra in the wavelength 
interval $\sim 200$ to $\sim 800$ nm for ice 
(solid curves and open symbols) and water (dashed curves and solid points). 

4. Absorption as a function of wave number in $\mu$m$^{-1}$ for laboratory-grown ice crystals 
(open circles) and for South Pole ice at level 3 (830 m; triangular symbols) and level 17 
(level 970 m; square symbols). The dotted curve is an exponential fit to data for lab ice at 
600 to 800 nm.

5. Comparison of absorption spectrum in ice and a number of other transparent solids. The 
straight lines representing an inverse fourth-power law, $\lambda^{-4}$, are meant to 
show that portions of each spectrum are due to Rayleigh scattering.

6. Wavelength dependence of the absorption on dust inferred from the
AMANDA laser measurements. The normalization is arbitrarily set to
unity at $\lambda$= 475 nm. Open squares, and the error bars, refer
to the model independent analysis where the intrinsic absorption in
ice has been subtracted. Triangles and circles are values 
from AMANDA depths 830 and 970 m, respectively, using an exponential fit to 
the intrinsic ice absorption. The band within the dashed lines 
corresponds to the wavelength dependence measured for atmospheric dust
at a US location \cite{15}.

7. Ratio of dust concentration at 970 m and 830 m as obtained by the 
analysis in this paper. Note that the ratio does not depend on wavelength,
which is a consistency check of the scheme employed.

8. The absorption length (i.e. the inverse of the absorption
coefficient) given by scattering on impurities in the ice, as a function
of wavelength for AMANDA 830 m level (filled triangles) and  
970 m (filled squares).
The values were obtained by subtracting the best fit exponential for
the intrinsic absorption from the measured absorption.

9. Three-component model of absorption by ice containing insoluble dust but 
no dissolved impurities. The concentrations of dust increase by
factors of 1.4 for each curve.

\newpage
\begin{center}
\begin{tabular}{|c|c|c|}
\hline Wavelength (nm) & $\lambda_{abs}^{830m}$ (m) & 
$\lambda_{abs}^{970m}$ (m)\\
\hline
\hline
410&$219.9\pm12.5$&$-$\\ \hline
415&$225.6\pm 6.2$& $168.7\pm 23.6$\\ \hline
420&$227.4\pm 4.0$& $163.9\pm 7.1$\\ \hline
435& $207.4\pm 1.5          $& $166.3\pm 1.9         $\\ \hline
450& $171.3\pm 1.3          $& $146.0\pm 2.1         $\\ \hline
475& $117.1\pm 0.7          $& $104.6\pm 1.2         $\\ \hline
500& $68.9\pm 1.3          $& $64.1\pm 1.0         $\\ \hline
530& $38.3 \pm 1.0          $& $39.5\pm 0.7         $\\ \hline
590& $12.0\pm 0.1          $& $10.5\pm 0.3         $\\ \hline
610&$10.3\pm 0.3          $&$-$\\ \hline
\hline
\end{tabular}
\end{center}
\vskip .1cm
\begin{itemize}
\item[Table 1.]
Absorption lengths $\lambda_{abs}$ 
(i.e. the inverse of the absorption coefficients $\alpha_{abs}$ defined in 
Eq.\,(~\ref{cabs})
for several laser wavelengths at two different depths, around 830 and 970 m,
determined by the AMANDA calibration system in the South Pole ice. For 410 and
610 nm, only data for 830 m could be obtained.

\end{itemize}
\end{document}